# Inducing the controlled rotation of single

# o-MeO-DMBI molecules anchored on Au(111)

*Frank Eisenhut,[1,2] Jörg Meyer,[1,2] Justus Krüger,[1,2] Robin Ohmann,[1,2,#] Gianaurelio Cuniberti,[1,2,3] and Francesca Moresco [2,1*]*

[1]Institute for Materials Science and Max Bergmann Center of Biomaterials, TU Dresden, 01062 Dresden, Germany

[2]Center for Advancing Electronics Dresden, TU Dresden, 01062 Dresden, Germany

[3]Dresden Center for Computational Materials Science (DCMS), TU Dresden, 01062 Dresden, Germany.

[*]Corresponding author: francesca.moresco@tu-dresden.de

[#]Present address: Energy Materials and Surface Sciences Unit (EMSS), Okinawa Institute of Science and Technology Graduate University (OIST), 1919-1 Tancha, Onna-son, Okinawa, 904-0495 Japan.

**Keywords:** Molecular rotor, scanning tunneling microscopy (STM), adsorption, manipulation, voltage pulses

**Abstract**

A key step towards building single molecule machines is to control the rotation of molecules and nanostructures step by step on a surface. Here, we used the tunneling electrons coming from the tip of a scanning tunneling microscope to achieve the controlled directed rotation of complex o-MeO-DMBI molecules. We studied the adsorption of single o-MeO-DMBI molecules on Au(111) by scanning tunneling microscopy at low temperature. The enantiomeric form of the molecule on the surface can be determined by imaging the molecule by STM at high bias voltage. We observed by lateral manipulation experiments that the molecules chemisorb on the surface and are anchored on Au(111) with an oxygen-gold bond *via* their methoxy-group. Driven by inelastic tunneling electrons, o-MeO-DMBI molecules can controllably rotate, stepwise and unidirectional, either clockwise or counterclockwise depending on their enantiomeric form.

**Highlights**

- Adsorption of single o-MeO-DMBI molecules on Au(111). The molecule is chemiadsorbed *via* the methoxy-group, resulting in a new robust molecular rotor.

- By lateral manipulation with the STM tip, the molecule can be rotated around its anchor point.

- Driven by tunneling electrons, the o-MeO-DMBI molecules can be rotated step by step either clockwise or counterclockwise in a controlled manner.

## 1. Introduction

Scanning tunneling microscopy (STM) does not only allow to image conducting surfaces and molecular adsorbates with high spatial resolution, but provides a nanoscale probe able to manipulate matter at the atomic scale [1]. The STM tip can be used to induce chemical reactions [2, 3], to translate [4], rotate [5], and switch single molecules [6, 7], or to drive nanocars [8-10]. A pioneer in the field of manipulation and the control of single atoms and molecules using an STM-tip was Karl-Heinz Rieder [11-13]. Also inspired by his ideas many researchers planned to construct molecular machines on-surfaces by using single molecules [14].

One key element for the construction of molecular machines are molecular motors and rotors. Several approaches for building single molecule rotors have been proposed in the last twenty years [5, 15-17]. Different excitation modes, as for example light, thermal excitation, mechanical forces, or tunneling electrons [18-20] and a few possibilities for anchoring the molecules on a surface have been investigated [21-23]. Strategies for the transmission of motion have been presented, showing examples of molecular rotors able to move and rotate single atoms or molecules [24, 25]. Moreover, methods to turn on and off molecular rotors have been demonstrated [21, 22]. However, the examples of unidirectional rotations are rare [18, 19, 25], while the selective triggering of the rotor in one direction to a predefined position is of great importance for the development of molecular machines.

In this paper, we show by STM imaging and manipulation that single o-MeO-DMBI molecules are anchored to the Au(111) surface by means of their methoxy-group. By exciting the molecule *via* tunneling electrons, we can controllably rotate the molecule into all possible adsorption orientations, clockwise or counterclockwise depending on the molecular chirality.

## 2. Experimental details

2-(2-Methoxyphenyl)-1,3-dimethyl-1H-benzoimidazol-3-ium (o-MeO-DMBI) molecules were studied on the Au(111) surface by STM at low temperature (T ≈ 5 K) in ultra-high vacuum (UHV) conditions (P ≈ $1 \times 10^{-10}$ mbar). STM measurements were performed in constant-current mode with the bias voltage applied to the sample. A Au(111) single crystal was used as substrate and prepared by repeated cycles of sputtering (Ar$^+$) and annealing (730 K). After this cleaning procedure, o-MeO-DMBI-I (synthesis described in [26]) was evaporated from a Knudsen cell heated at a temperature of 490 K. Subsequently, the sample was cooled down and transferred into the STM. To ensure that no manipulation happens during topographic imaging, the STM images were normally recorded at non-perturbative conditions (I ≤ 100 pA, V ≤ 0.5 V). Lateral manipulation in constant current mode with a tunnel resistance R ≈ $10^6$ Ω was used to identify the anchoring point of the molecule. Furthermore, the controlled rotation of the molecules was induced by tunneling electrons, precisely positioning the STM tip on the DMBI-molecule and applying a voltage pulse.

## 3.    Results and discussion

### 3.1    Adsorption of o-MeO-DMBI on the Au(111) surface

We deposited the o-MeO-DMBI-I molecular precursors at submonolayer coverage on a clean Au(111) surface kept at room temperature. During evaporation at 490 K, the precursors are reduced by cleaving iodine and resulting in o-MeO-DMBI (Figure 1a). Due to this reduction, that has been verified by X-ray photoelectron spectroscopy elsewhere [26], the sublimated molecules work as n-dopant in organic electronic devices and were widely investigated for applications in organic electronics [27, 28]. The iodine atoms adsorb on the Au(111) surface and are visible as small round protrusions in the STM images, as shown in Figure 1b. Caused by the iodine adsorption, the surface restructures and the distances between the next-nearest neighbor soliton walls increases by about 1.5 nm as compared to the unaltered herringbone reconstruction, as known in the literature [29, 30].

As one can see in Figure 1b, several isolated molecules are visible in the STM image, and appear as elongated protrusions superposed to a more intense circular feature when imaged at low voltage. In contrast to the non-symmetric structure of the molecule due to the methoxy-group (Figure 1a), a clear mirror symmetry is present in the STM image at a low bias voltage.

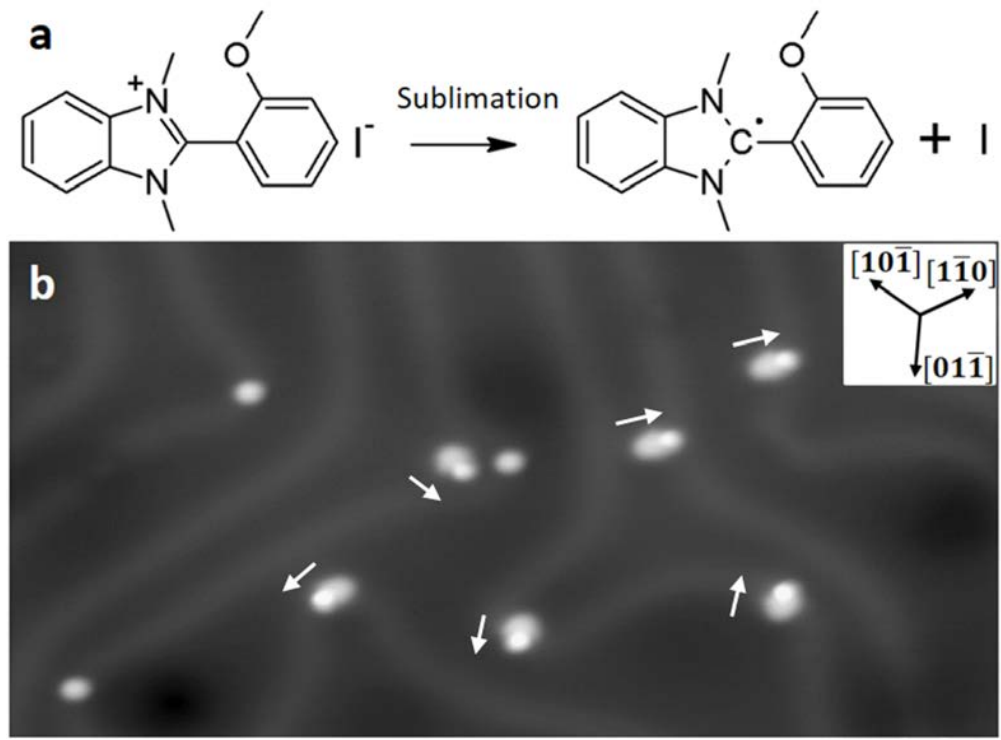

**Figure 1**. *o-MeO-DMBI molecules sublimated on Au(111). (a) Chemical structures of the precursor molecule o-MeO-DMBI-I that is reduced after thermal sublimation by iodine and forms the molecule o-MeO-DMBI. (b) Overview STM image of a submonolayer of o-MeO-DMBI on Au(111). The small circular protrusions can be assigned to iodine atoms. The arrows next to the molecules show the adsorption orientations of the corresponding molecule on Au(111). The inset shows the main crystallographic directions of the surface (image size: 40 nm × 20 nm; V = -0.4 V; I = 60 pA).*

The molecules preferably adsorb either at the elbow site or at the soliton walls of the Au(111) herringbone reconstruction [31, 32] and their orientation is determined by the surface structure. O-MeO-DMBI shows six preferred adsorption orientations on Au(111) (at a relative angle of 60°) corresponding to the high symmetry directions of the hexagonal surface lattice (see inset in Figure 1b). Five of the six possible preferred orientations are visible in Figure 1b and highlighted by arrows next to the molecules.

### 3.2    Lateral manipulation and anchoring point

The presence of a methoxy-group and of a possible radical form (Figure 1a) suggests a strong molecule-surface interaction that can be tested by pushing the single molecules by lateral manipulation with the STM tip.

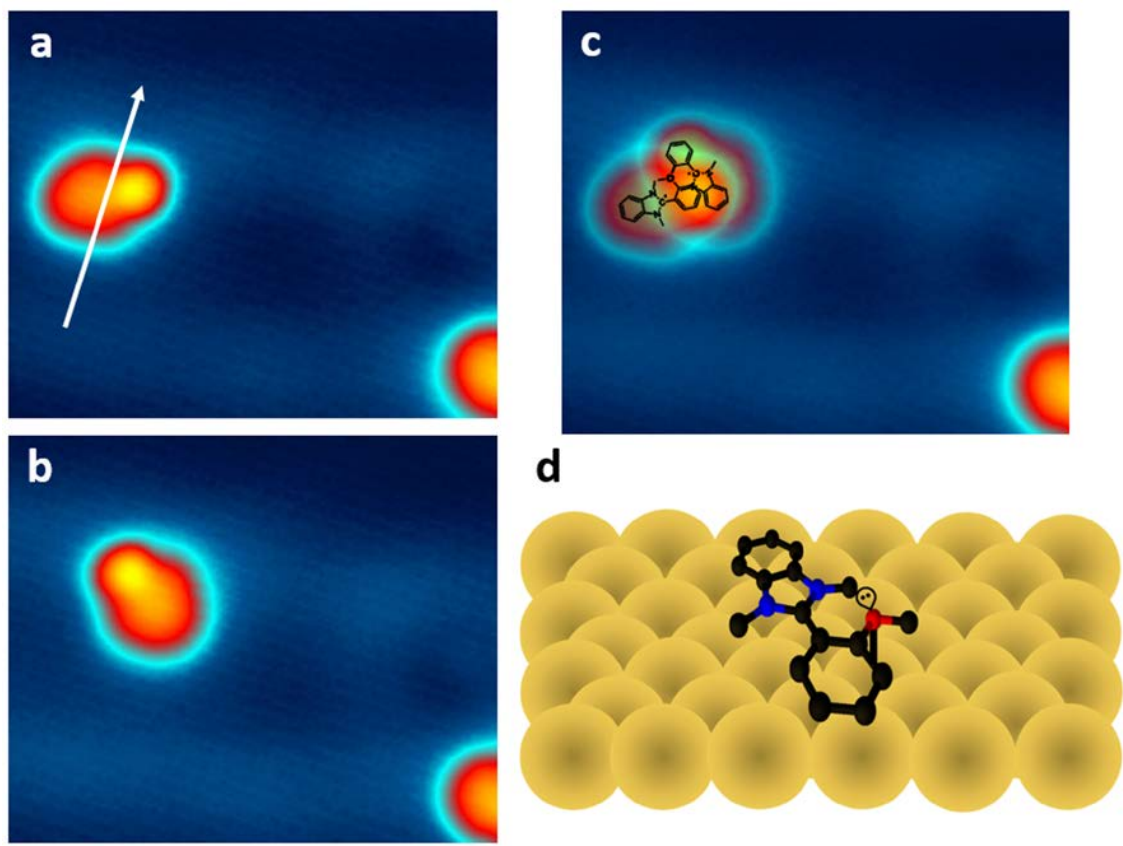

***Figure 2****. Lateral manipulation and adsorption of a o-MeO-DMBI molecule. (a) STM image of one molecule on Au(111). Lateral manipulation is performed along the white arrow (parameters: I = 4.0 nA; V = 10 mV). (b) STM image of the same area after lateral manipulation showing the changed orientation of the molecules. (image size 6 nm × 5 nm; V = 0.2 V; I = 30 pA). (c) Superimposed image of (a) and (b). By inserting the molecular structures into the superimposed image, the group anchoring the molecule to the surface can be identified as the methoxy-group. (d) Scheme of the adsorption of o-MeO-DMBI on Au(111).*

During lateral manipulation, the bias voltage is kept very low (V ~ 10 mV) and the tip is moved over the molecule at low tip-surface distance [1, 4]. One typical example of lateral manipulation is shown in Figure 2. Starting from the image of Figure 2a, the STM tip is moved along the white arrows at manipulation conditions that would normally result in the translation of purely physisorbed molecules [33, 34]. As one can see in Figure 2b, the molecules do not translate, but change the orientation, rotating around a fixed anchoring point. Manipulation trials with reduced tip height and consequently stronger tip-molecule interaction never lead to a translation of the molecule. Further tip height reduction during manipulation results in picking up the molecule together with one of the underlying gold atoms, imaged afterwards as a hole in the surface (see Figure S1). The lateral manipulation experiments demonstrate that the molecules are not physisorbed, but are chemically anchored to the Au(111) surface, as expected considering the similar case of thioether-molecules, which are known to strongly bind to surface atoms on coinage metals [35]. We therefore conclude that the o-MeO-DMBI molecule is anchored to a gold atom through one of the two lone pairs of electrons of the oxygen (Figure 2d). To confirm this conclusion and identify the anchoring point (assigned to the oxygen) in the STM images, we superimpose in Figure 2c the STM image of the molecule before (Figure 2a) and after (Figure 2b) the lateral manipulation, with the corresponding molecular structure. No anchoring was observed at the radical position of the molecule, as the radical is probably passivated by a hydrogen atom upon sublimation or adsorption [36].

### 3.3 Molecular chirality on the Au(111) surface

At higher bias voltage, the STM image of a single O-MeO-DMBI molecule is not symmetric anymore (Figure 3) and show two chiral forms on the surface (Figure 3a-b and Figure 3c-d). One can clearly observe the mirror symmetry between the images of Figure 3a and 3c. By superposing the molecular structure to both chiralities (Figure 3b and 3d) we can confirm the position of the central anchoring oxygen atom and assign a surface chirality to the molecules. We denote the two chiral forms as "R" and "S" enantiomer by the Cahn-Ingold-Prelog rules [18]. The "R"-enantiomer is thereby defined with the oxygen-atom as center. Methyl group, gold atom and phenyl ring are oriented clockwise (Figure 2d). For the "S"-enantiomer these groups are oriented counterclockwise. As a consequence, we can also assign the methoxyphenyl-part of the molecule to the more intense circular feature in the STM images recorded at lower voltage (see Figure 1).

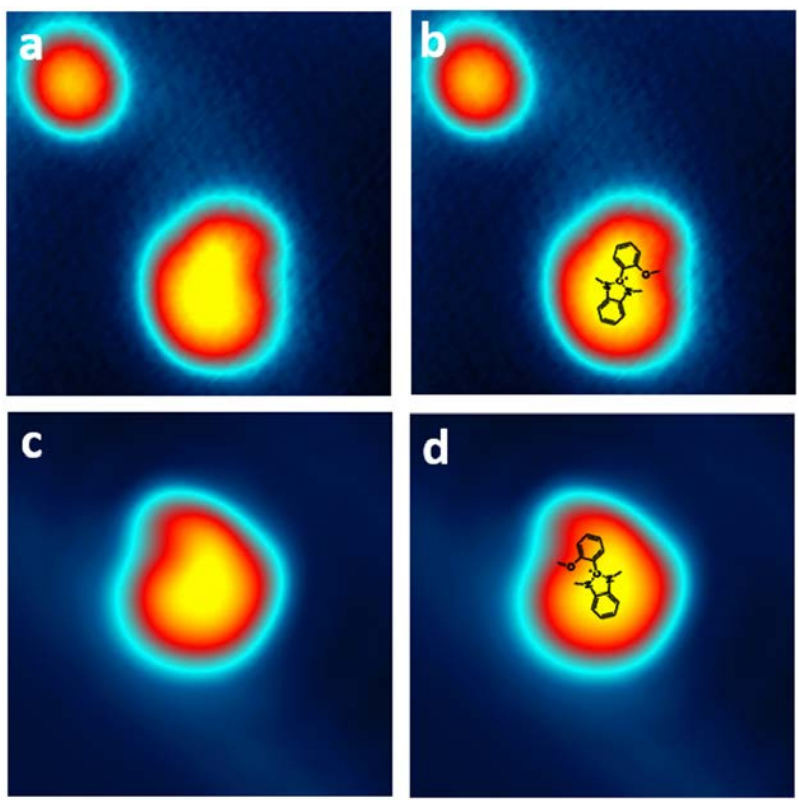

***Figure 3.*** *Single o-MeO-DMBI molecules imaged at higher bias-voltage showing the two different enantiomeric forms. (a) S-enantiomer, (b) same as (a) with superposed molecular structure. (c) R-enantiomer showing a mirror symmetry in comparison to the molecule in (a). (d) Same STM image as in (c) with inserted molecular structure. (I = 25 pA; V = 2.5 V; image sizes: 4.0 nm × 4.0 nm).*

### 3.4    Manipulation by tunneling electrons and molecular rotor

By increasing the bias voltage over 2.5 V, the molecules start to appear wheel-like (Figure 4). This is due to the fast rotation of the molecules during imaging. In these conditions, a superposition of the molecules in their six equivalently preferred orientations on the Au(111) surface is visible in the STM images (Figure 4b). This suggests that tunneling electrons can possibly induce the rotation of the molecule. To stop the rotation during imaging, we imaged the same area at a slightly lower bias voltage (2.5 V instead of 2.7 V) and lower current (Figure 4c). At these conditions, the upper molecule still rotates, but with a lower frequency. The lower molecule is chemisorbed on the soliton wall and remains therefore unchanged. This fact shows that the potential energy surface for the molecule is influenced by the adsorption position on the herringbone reconstruction of the Au(111) surface [37, 38]. In this case, please note that the molecule corresponds to an "S"-enantiomer.

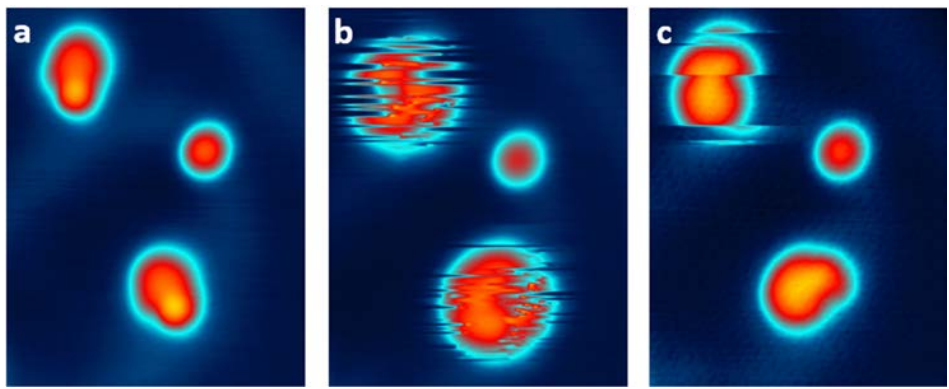

**Figure 4.** *o-MeO-DMBI molecules imaged at different bias-voltages. (a) STM image of two o-MeO-DMBI molecules at non-perturbative imaging conditions (I = 100 pA; V = 0.2 V) appearing symmetric. (b) Same area as (a) imaged at a higher voltage (I = 100 pA; V = 2.7 V). The molecules appear as wheels due to their rotation. (c) Same area as (a) imaged at high voltage and low current (I = 25 pA; V = 2.5 V); the upper molecule rotates slowly, while the other is not rotating and appears non-symmetric due to the methoxy-group (images size: 5.5 nm × 7.0 nm).*

After observing that the molecules are rotating during imaging at high voltage, we investigated the controlled rotation by voltage pulses. As shown in Figure 5, we positioned the tip above a single molecule at the positions marked with the cross (Figure 5a) and increased the voltage. In this example, the feedback-loop was kept closed and the voltage ramped from 0.2 V to 3.0 V.

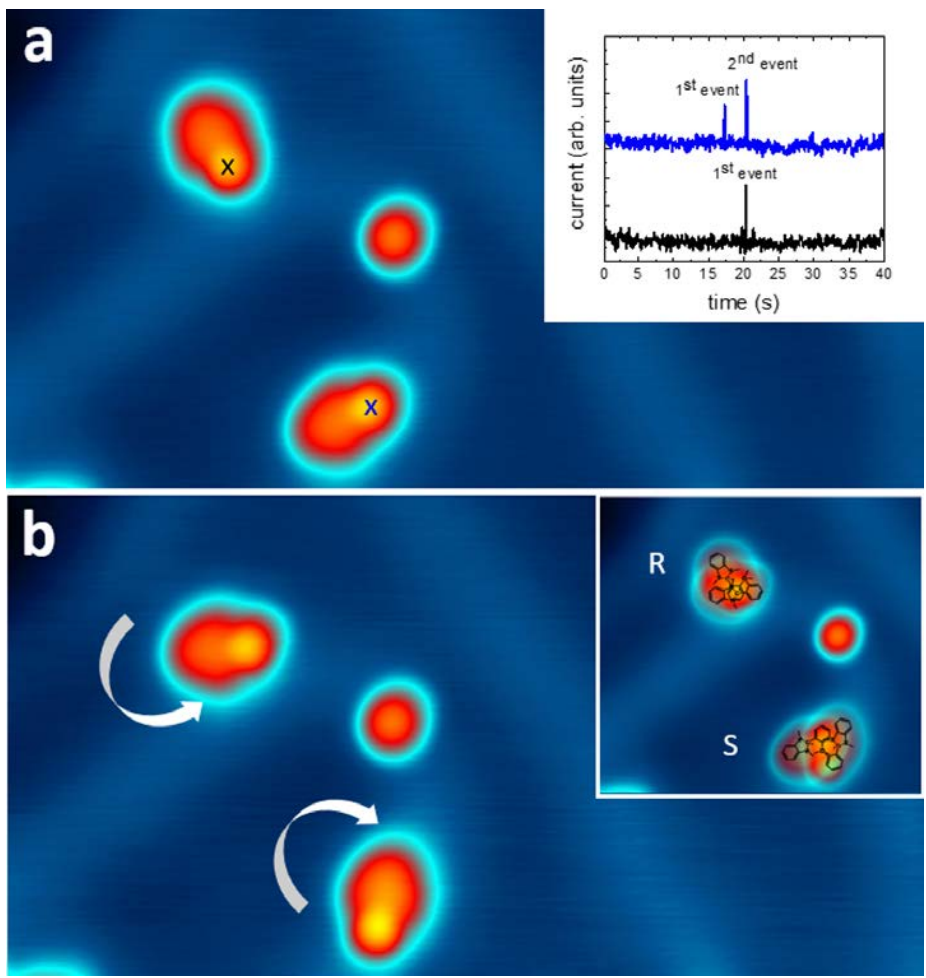

**Figure 5.** *Dependence of the rotation direction for different enantiomeric o-MeO-DMBI molecules induced by voltage pulses. (a) Image of two o-MeO-DMBI molecules. The crosses are marking the positions of the applied voltage pulses. (voltage pulse parameter: closed feedback loop; I = 39 pA; voltage ramped from 0.2 V to 3.0 V; t = 40 s) In the inset the current traces of the voltage pulses with the corresponding color in (a) are shown. Rotational events can be identified as peaks in the current traces. (b) The same area as in (a) after the voltage pulses. The upper molecule rotated counterclockwise, while the lower one rotated clockwise. The inset in (b) shows the superposed image of (a) and (b). Considering the superposed molecular structures, one can confirm that the upper molecule is a "R"-enantiomeric form and the lower molecule a "S"-enantiomeric form, both rotating around the O-Au bond. (image size: 12.5 nm × 6.5 nm; V = -0.2 V; I = 170 pA).*

Due to the reaction time of the feedback-loop, one can observe peaks in the current traces corresponding to rotational events as shown in the inset of Figure 5a. In the case of the upper molecule, the voltage ramp (black curve) lead to one peak, while for the bottom molecule two peaks are present in the current trace. We therefore expect that the upper molecule rotated once to the nearest-neighbor equivalent orientation (hopping by ± 60°) and the lower molecule rotated twice, reaching the next-nearest neighbor orientation (hopping by ± 120°). Indeed, this is exactly what we observed imaging the same area once more (Figure 5b). The upper molecule rotated one step counterclockwise (-60°), while the bottom molecule rotated two steps in the clockwise direction (+120°).

In this case, the chirality of the two molecules is determined by superimposing images 5a and 5b together with the inserted molecular structure (inset in Figure 5b and enlarged in Figure S2 of the supporting information). Considering that the upper molecule is in the "R"-enantiomeric form and the lower molecule is in the "S"-enantiomeric form, we observed that the "S"-enantiomer rotates clockwise and "R"-enantiomer counterclockwise. After testing dozens of molecules, we could conclude that the chirality of the molecules in all observed cases determines the direction of rotation, defining whether a molecule rotates clockwise ("S"-enantiomer) or counterclockwise ("R"-enantiomer), as observed also in other cases [35, 39, 40].

We further investigated the rotation applying hundreds of voltage pulses. We could detect the first rotation events at voltages of about 0.4 V. This is in good agreement with previous publications proposing that in this voltage range the tunnel electrons excite a C-H vibrational mode [16, 41], able to induce the rotation of the molecule [18]. Furthermore, we could exclude electric field effects by increasing the bias voltage to 4.0 eV and retracting the tip by 1.5 nm. In these conditions we never observed rotational events.

Finally, we demonstrated the complete step by step rotation of a molecule through all six possible equivalent adsorption orientations (Figure 6). To this aim, we optimized the voltage pulse parameters to obtain a one-step jump in a reasonable and controllable time (of the order of a second) into the nearest neighbor orientation (± 60°).

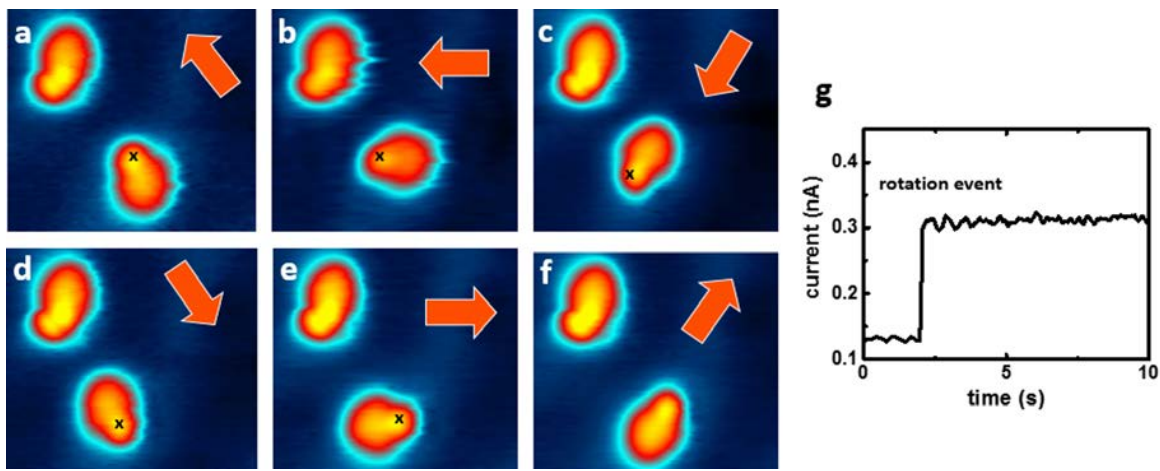

***Figure 6***. *Controlled counterclockwise stepwise rotation of an o-MeO-DMBI molecule. (a) – (f) STM images of two o-MeO-DMBI molecules. The molecule at the bottom was rotated counterclockwise step by step into all possible orientations via voltage pulses applied at the position marked with the black cross. (Manipulation parameters: V = 0.7 V and I = 0.1 nA up to 0.5 nA, open feedback-loop; Images size: 5.0 nm × 4.5 nm; Image conditions: V = 0.5 V and I = 80 pA). For clarity, the adsorption orientation of the molecule is marked with an orange arrow. (g) Current signal recorded during manipulation in (a): A single rotation event can be observed by a jump in the current trace.*

An open feedback-loop at 0.7 V and about 100 pA was used and all tested molecules rotated unidirectional, clockwise or counterclockwise depending on their enantiomeric form. Exemplary, we rotated an o-MeO-DMBI molecule in the "R"-enantiomeric form (molecule at the bottom of Figure 6a). Using this approach, all six preferred orientations could be reached step by step counterclockwise by applying voltage pulses at the positions marked with the black cross (see Figure 6a to Figure 6f), in agreement with the case of Figure 5. The current height trace for one rotational event is presented in Figure 6g (recorded during the pulse of Figure 6a).

## 4.    Conclusion

In summary, we presented the adsorption of single o-MeO-DMBI molecules on Au(111). The molecules chemisorb on the surface by means of a localized oxygen-gold bond. The position of the anchor methoxy-group and thus the enantiomeric form can be determined by imaging the molecule by STM at high bias voltage. Depending on the molecular chirality, o-MeO-DMBI molecules can controllably rotate either clockwise ("S"-enantiomer) or counterclockwise ("R"-enantiomer) by means of voltage pulses. The presented method can be used to systematically manipulate a molecular rotor into a desired orientation on the surface. Our approach shows that complex molecules can be rotated on coinage metal surfaces by inducing methoxy-groups in the chemical structure, opening new routes for the construction of complex molecule machines.

**Acknowledgements**

This work has received funding from the European Union's Horizon 2020 research and innovation programme under the project MEMO, grant agreement No 766864.

Support by the German Excellence Initiative *via* the Cluster of Excellence EXC1056 ''Center for Advancing Electronics Dresden'' (cfaed), the International Helmholtz Research School ''NANONET'' and the DFG and the National Science Foundation *via* the common project NSF 11-568 is gratefully acknowledged.

The authors thank Benjamin Naab and Zhenan Bao for providing the o-MeO-DMBI-I molecules, and Christian Joachim for fruitful discussion.

**Abbreviations**

o-MeO-DMBI-I – 2-(2-Methoxyphenyl)-1,3-dimethyl-1H-benzoimidazol-3-ium iodide

o-MeO-DMBI – 2-(2-Methoxyphenyl)-1,3-dimethyl-1H-benzoimidazol-3-ium

DMBI –1,3-dimethyl-1H-benzoimidazol-3-ium

Graphical abstract

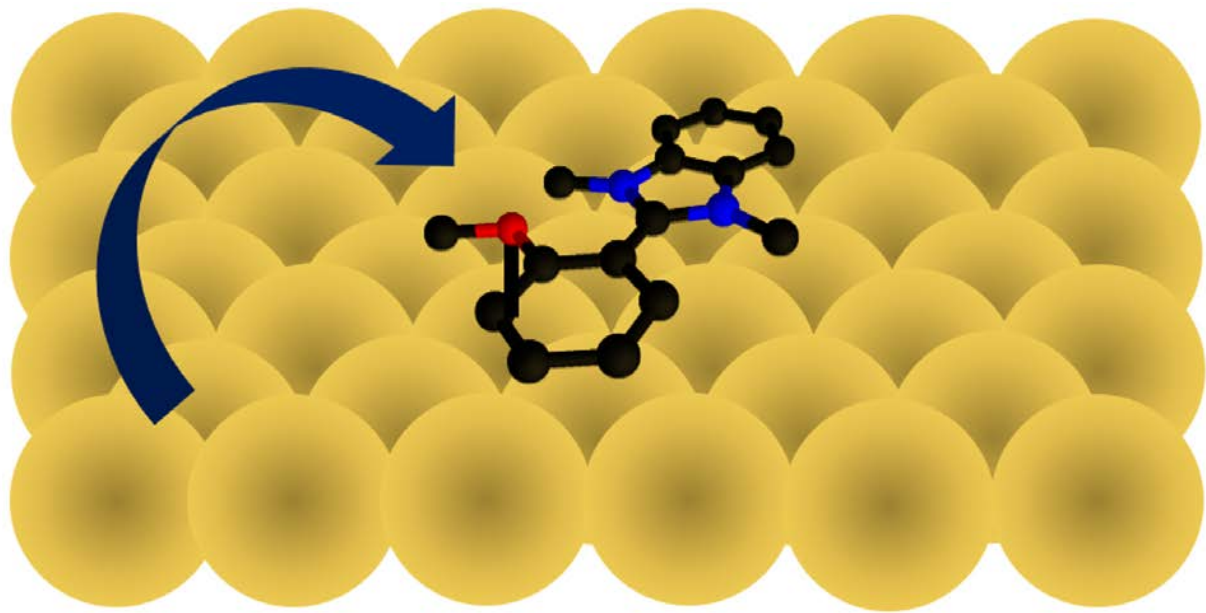